\documentclass[12pt]{article}
\newtheorem{remark}{Remark}

\newfont{\BB}{msbm10 scaled\magstep1}

\def\D{\mbox{\BB D}}
\def\R{\mbox{\BB R}} 

\begin{document}

\title{Quantum Schr\"{o}dinger bridges\\{\small {\em Dedicated to {\em
Anders Lindquist} on the occasion of his 60th birthday}}}

\author{M. Pavon\\
Dipartimento di Matematica Pura e Applicata,\\
Universit\`a di Padova,\\
via Belzoni 7,\\
and LADSEB-CNR,\\
35131 Padova,\\
Italy.\\
{\tt pavon@math.unipd.it}}
\maketitle

\section*{Abstract}
Elaborating on M. Pavon, {\em J.
Math. Phys.} {\bf 40} (1999), 5565-5577, we develop a simplified version of a
variational principle within Nelson stochastic mechanics that
produces the von Neumann
wave packet reduction after a position measurement. This stochastic
control problem
parallels, with a different kinematics, the problem of the
Schr\"{o}dinger bridge. This gives a profound meaning  to what was
observed by Schr\"{o}dinger in 1931 concerning Schr\"{o}dinger bridges:
``{\em Merkw\"{u}rdige
Analogien zur Quantenmechanik, die mir sehr des Hindenkens wert erscheinen}".

\section{Introduction: Schr\"{o}dinger's problem}

In 1931/32 \cite{S,S2}, Schr\"{o}dinger considered the following problem.
A cloud of $N$ Brownian particles in $\R^n$ has been
observed having at time $t_0$ an empirical distribution approximately
equal to
$\rho_0(x)dx$. At some later time $t_1$, an empirical
distribution approximately equal to  $\rho_1(x)dx$ is observed.
Suppose that $\rho_1(x)$ considerably differs from what it should
be according to the law of large numbers ($N$ is large), namely
$$\int_{t_0}^{t_1}p(t_0,y,t_1,x)\rho_0(y)dy,
$$

where
$$p(s,y,t,x)=\left[2\pi(t-s)\right]
^{-\frac{n}{2}}\exp\left[-\frac{|x-y|^2} {2(t-s)}\right],\quad s<t,
$$
is the transition density of the Wiener process. It is apparent that the
particles have been transported in an unlikely way. But of the
many unlikely ways in which this could have happened, which one is
the most likely?

In modern probabilistic language, this is a problem of large deviations of
the empirical distribution \cite{F2}. By discretization and passage to the
limit, Schr\"{o}dinger computed the most likely intermediate empirical
distribution as $N\rightarrow \infty$. It turned out that the optimal
random evolution, the {\em Schr\"{o}dinger bridge} from $\rho_0$ to
$\rho_1$ over Brownian motion, had at each time a density
$\rho(\cdot,t)$ that factored as $\rho(x,t)=\phi(x,t)\hat{\phi}(x,t)$,
where $\phi$ and $\hat{\phi}$ are a $p$-harmonic and a $p$-coharmonic
functions, respectively. That is
\begin{eqnarray}\label{SY1}
&&\phi(t,x)=\int
p(t,x,t_1,y)\phi(t_1,y)dy,\\&&\hat{\phi}(t,x)=\int
p(t_0,y,t,x)\hat{\phi}(t_0,y)dy.\label{SY2}
\end{eqnarray}
The existence and uniqueness of a pair $(\phi,\hat{\phi})$ satisfying
(\ref{SY1})-(\ref{SY2}) and the boundary conditions
$\phi(x,t_0)\hat{\phi}(x,t_0)=\rho_0(x)$,
$\phi(x,t_1)\hat{\phi}(x,t_1)=\rho_1(x)$ was guessed by Schr\"{o}dinger
on the basis of his intuition. He was later shown to be quite right in
various degrees of generality by Fortet \cite{For}, Beurlin \cite{Beu},
Jamison \cite{Jam}, F\"{o}llmer \cite{F2}. Jamison showed, in particular,
that the Schr\"{o}dinger bridge is the unique {\em Markov} process
$\{x(t)\}$ in the class of {\em reciprocal processes} (one-dimensional
Markov fields) introduced by Bernstein \cite{BER} having as two-sided
transition density
$$q(s,x;t,y;u,z)=\frac{p(s,x;t,y)p(t,y;u,z)}{p(s,x;u,z)},\quad s<t<u,
$$
namely $q(s,x;t,y;u,z)dy$ is the probability of finding the process $x$ in
the volume $dy$ at time $t$ given that $x(s)=x$ and $x(u)=z$.
Schr\"{o}dinger was struck by the following remarkable property of the
solution: The Schr\"{o}dinger bridge from
$\rho_1$ to
$\rho_0$ over Brownian motion is just the time reversal of the
Schr\"{o}dinger bridge from $\rho_0$ to
$\rho_1$. In Schr\"{o}dinger's words: ``Abnormal states have arisen with
high probability by an exact time reversal of a proper diffusion
process". This led him to entitle \cite{S}: ``On the reversal of
natural laws" A few years later, Kolmogorov wrote a paper on the subject
with a very similar title \cite{Kol}. Moreover, the fact that the
Schr\"{o}dinger bridge has density $\rho(x,t)=\phi(x,t)\hat{\phi}(x,t)$
resembles the fact that in quantum mechanics the density may be expressed
as $\rho(x,t)=\psi(x,t)\bar{\psi}(x,t)$. This analogy has inspired various
attempts to construct a stochastic reformulation of quantum mechanics
\cite{Z}-\cite{LK2} starting from \cite{S, S2,BER}. In order to
discuss a more general
Schr\"{o}dinger bridge problem, we recall in the next session some
essential facts  on
the kinematics of finite-energy diffusions as presented in \cite{N1,F,N3,KS}.

\section{Elements of Nelson-F\"{o}llmer kinematics of finite energy
diffusions}

Let $(\Omega,{\cal F},{\bf P})$ be a complete probability space. A stochastic
process $\{\xi(t);t_0\le t\le t_1\}$ mapping $[t_0,t_1]$ into
$L^2_n(\Omega,{\cal F},{\bf P})$ is called a {\it finite-energy
diffusion} with constant diffusion coefficient $I_n\sigma^2$ if the path
$\xi(\omega)$ belongs a.s. to $C([t_0,t_1];\R^n)$ (n-dimensional
continuous functions) and

\begin{equation}\label{K1}
\xi(t)-\xi(s)=\int_s^t\beta(\tau)d\tau+\sigma w_+(s,t),\quad
t_0\le s<t\le t_1, \end{equation}
where the {\it forward drift} $\beta(t)$ is at each time
$t$ a measurable function of the past $\{\xi(\tau);0\le \tau\le t\}$, and
$w_+(\cdot,\cdot)$ is a standard, n-dimensional {\it Wiener difference
process} with the property that
$w_+(s,t)$ is independent of
$\{\xi(\tau);0\le \tau\le s\}$. Moreover, $\beta$ must satisfy the
finite-energy condition
\begin{equation}\label{K2}
E\left\{\int_{t_0}^{t_1}\beta(\tau)\cdot\beta(\tau)d\tau\right\}<\infty.
\end{equation}
We recall the characterizing properties of the n-dimensional
{\it Wiener difference process} $w_+(s,t)$, see \cite[Chapter 11]{N1} and
\cite[Section 1]{N3}. It is a process such that $w_+(t,s)=-w_+(s,t)$,
$w_+(s,u)+w_+(u,t)=w_+(s,t)$, and that $w_+(s,t)$ is Gaussian
distributed with mean
zero and variance $I_n|s-t|$. Moreover, (the components of) $w_+(s,t)$ and
$w_+(u,v)$ are independent whenever $[s,t]$ and $[u,v]$ don't overlap. Of
course, $w_+(t):=w_+(t_0,t)$ is a standard Wiener process such that
$w_+(s,t)=w_+(t)-w_+(s)$.
In \cite{F}, F\"{o}llmer has shown that a finite-energy diffusion also
admits a reverse-time differential. Namely, there exists a measurable
function $\gamma(t)$ of the future $\{\xi(\tau);t\le \tau\le t_1\}$ called
{\it backward drift}, and another Wiener difference process
$w_-$ such that
\begin{equation}\label{K3}
\xi(t)-\xi(s)=\int_s^t\gamma(\tau)d\tau+\sigma w_-(s,t),\quad t_0\le
s<t\le t_1.
\end{equation}
Moreover, $\gamma$ satisfies
\begin{equation}\label{K4}
E\left\{\int_{t_0}^{t_1}\gamma(\tau)\cdot\gamma(\tau)d\tau\right\}<\infty,
\end{equation}
and
$w_-(s,t)$ is independent of $\{\xi(\tau);t\le \tau\le t_1\}$. Let us
agree that
$dt$ always indicates a strictly positive variable. For any function
$f$ defined on $[t_0,t_1]$,
let
    $$d_+f(t)=f(t+dt)-f(t)$$ be the {\it forward increment} at time $t$, and
$$d_-f(t)=f(t)-f(t-dt)$$ be the {\it backward increment} at time $t$.
For a finite-energy
diffusion, F\"{o}llmer has also shown in \cite{F} that the forward
and backward drifts may be
obtained as Nelson's conditional derivatives, namely
$$\beta(t)=\lim_{dt\searrow
0}E\left\{\frac{d_+\xi(t)}{dt}|\xi(\tau), t_0 \le
\tau \le t\right\},$$ and $$\gamma(t)=\lim_{dt\searrow
0}E\left\{\frac{d_-\xi(t)}{dt}|\xi(\tau), t
\le  \tau \le t_1\right\},$$
the limits being taken in $L^2_n(\Omega,{\cal F},P)$. It was finally
shown in \cite{F}
that the one-time probability density $\rho(\cdot,t)$ of $\xi(t)$
(which exists for every
$t>t_0$) is absolutely continuous on
$\R^n$ and the following duality relation holds $\forall t>0$
\begin{equation}\label{K4'}
E\{\beta(t)-\gamma(t)|\xi(t)\} =
\sigma^2\nabla\log\rho(\xi(t),t),\quad {\rm a.s.}
\end{equation}
\begin{remark} {\em It should be observed that in the study of
reverse-time differentials of diffusion processes, initiated by Nelson in
\cite{N00} and Nagasawa in
\cite{Na}, see \cite{Mo,HP} and references therein, important results
have obtained by A.
Linquist and G. Picci in the Gaussian case \cite{LP, LP2} {\em without
assumptions on the
reverse-time differential}. In particular, their results on
Gauss-Markov processes have
been crucial in order to develop a strong form of {\em stochastic
realization theory} \cite{LP}-\cite{LP4} together with a variety of
applications
\cite{LP4}-\cite{LM}.}
\end{remark}
Corresponding to (\ref{K1}) and (\ref{K3}) are two change of
variables formulas. Let
$f:\R^n\times [0,T] \rightarrow \R$ be twice continuously differentiable with
respect to the spatial variable and once with respect to time. Then,
if $\xi$ is a
finite-energy diffusion satisfying (\ref{K1}) and (\ref{K3}), we have
\begin{eqnarray}\nonumber
f(\xi(t),t)-f(\xi(s),s)=\int_s^t\left(\frac{\partial}{\partial
\tau}+\beta(\tau)\cdot\nabla+\frac{\sigma^2}{2}\Delta\right)f(\xi(\tau
),\tau)d\tau\\+\int_s^t\sigma\nabla
f(\xi(\tau),\tau)\cdot
d_+w_+(\tau),\label{K5}\\f(\xi(t),t)-f(\xi(s),s)=\nonumber
\int_s^t\left(\frac{\partial}{\partial
\tau}+\gamma(\tau)\cdot\nabla-\frac{\sigma^2}{2}\Delta\right)f(\xi(\tau),\tau)d\tau
\\+\int_s^t\sigma\nabla
f(\xi(\tau),\tau)\cdot d_-w_-(\tau). \label{K6} \end{eqnarray}
The stochastic integrals
appearing in (\ref{K5}) and (\ref{K6}) are a (forward) Ito integral
and a backward Ito integral,
respectively, see \cite {N3} for the details.

\section{Schr\"{o}dinger bridges}\label{SB}

The solution to the Schr\"{o}dinger problem can be obtained by
solving a stochastic control problem. The Kullback-Leibler
pseudo-distance between two
probability densities
$p(\cdot)$ and $q(\cdot)$ is defined by

$$H(p,q):=\int_{\R^n}\log\frac{p(x)}{q(x)}p(x)dx.$$
This concept can be considerably generalized. Let $\Omega:={\cal
C}([t_0,t_1],\R^n)$ denote the family of $n$-dimensional
continuous functions, let
$W_x$ denote Wiener measure on $\Omega$ starting at $x$, and let
$$W:=\int W_x\,dx
$$
be stationary Wiener measure. Let $\D$ be the family of
distributions on $\Omega$ that are equivalent to $W$. For
$Q,P\in\D$, we define  the {\it relative
entropy } $H(Q,P)$ of $Q$ with respect to $P$ as
$$H(Q,P)=E_Q[\log\frac{dQ}{dP}].$$
It then follows from Girsanov's theorem that
\cite{KS,F,F2}
\begin{eqnarray}\label{GIR1}
H(Q,P)=H(q(t_0),p(t_0))+E_Q\left[\int_{t_0}^{t_1}\frac{1}{2}
[\beta^Q(t)-\beta^P(t)]\cdot
[\beta^Q(t)-\beta^P(t)]dt\right]\nonumber\\=H(q(t_1),p(t_1))+E_Q
\left[\int_{t_0}^{t_1}\frac{1}{2}
[\gamma^Q(t)-\gamma^P(t)]\cdot
[\gamma^Q(t)-\gamma^P(t)]dt\right].\label{GIR2}
\end{eqnarray}
Here $q(t_0)$ is the marginal density of $Q$ at $t_0$, $\beta^Q$ and
$\gamma^Q$ are the
forward and the backward drifts of $Q$, respectively. Now let $\rho_0$ and
$\rho_1$ be two
everywhere positive probability densities. Let $\D(\rho_0,\rho_1)$
denote the set of distributions in $\D$ having the prescribed marginal
densities at $t_0$
and $t_1$. Given $P\in\D$, we consider the following problem:

$${\rm Minimize}\quad H(Q,P) \quad {\rm over} \quad \D(\rho_0,\rho_1).$$
In view of (\ref{GIR2}), this is a stochastic control
problem. It is connected through Sanov's theorem
\cite{F2,Wak} to a problem of large deviations  of the empirical
distribution, according to Schr\"{o}dinger original motivation. Namely, if
$X^1,X^2,\ldots$ is an i.i.d. sequence of random elements on $\Omega$
with distribution $P$, then the sequence
$P^n[\frac{1}{n}\sum_{i=1}^n\delta_{X^i}\in\cdot]$ satisfies a large
deviation principle with {\em rate function} $H(\cdot,P)$. \\
\noindent
If there is
at least one
$Q$ in
$\D(\rho_0,\rho_1)$ such that
$H(Q,P)<\infty$,
it may be shown that there exists a unique minimizer $Q^*$ in
$\D(\rho_0,\rho_1)$ called
{\em the Schr\"{o}dinger bridge} from $\rho_0$ to $\rho_1$ over $P$. If
(the coordinate
process under) P is Markovian with forward drift field $b_+^P(x,t)$ and
transition density
$p(\sigma,x,\tau,y)$, then $Q^*$ is also Markovian with forward drift field
$$b_+^{Q^*}(x,t)=b_+^P(x,t)+\nabla\log\phi(x,t),$$ where the
(everywhere positive) function
$\phi$ solves together
with another function $\hat{\phi}$ the system
(\ref{SY1})-(\ref{SY2}) with boundary conditions
$$\phi(x,t_0)\hat{\phi}(x,t_0)=\rho_0(x),\quad
\phi(x,t_1)\hat{\phi}(x,t_1)=\rho_1(x). $$
Moreover,  $\rho(x,t)=\phi(x,t)\hat{\phi}(x,t), \forall t\in [t_0,t_1]$.
This result has been suitably extended to the case where $P$ is
non-Markovian in \cite{PReg}.
For a survey of the theory of Schr\"{o}dinger bridges with an
extended bibliography see \cite{Wak}.
 
\vspace{0.2cm}
Consider now the following simpler problem: We have a {\em reference
stochastic model} $P\in \D$. We think of $P$ as modeling the macroscopic
evolution of a thermodynamic system. Suppose we observe at time
$t_1$ the (everywhere positive) density $\rho_1$ different from the
marginal density of $P$. Thus we need to solve the following optimization
problem
$${\rm Minimize}\quad H(Q,P) \quad {\rm over} \quad Q\in\D(\rho_1).$$
where $\D(\rho_1)$ denotes the set of distributions in $\D$ having
density $\rho_1$ at $t_1$. Let us assume that $H(\rho_1,p(t_1))<\infty$.
In view of (\ref{GIR2}), this stochastic control problem can be trivially
solved. The unique solution is given
by the distribution
$Q^*$ having backward drift
$\gamma^P(t)$ and marginal density $\rho_1$ at time $t_1$. Thus, the
result of measurement at time $t_1$ leads to the replacement of the
stochastic model $P$ with $Q^*$. Notice that the backward drift
$\gamma^P(t)$ is perfectly preserved by this procedure. Symmetrically, if
we were to change the initial distribution at time $t_0$, the procedure
would preserve the forward drift $\beta^P(t)$.

\section{Elements of Nelson's stochastic mechanics}
Nelson's stochastic mechanics is a quantization procedure for classical
dynamical systems based on diffusion processes. Following some early 
work by Feynes
\cite{Fe} and others, Nelson and Guerra elaborated a clean formulation
starting from 1966
\cite{N0,N1,G}, showing that the Schr\"{o}dinger equation could be derived
from a continuity type equation plus a Newton type law, provided one
accepted a certain  definition for the stochastic acceleration. In
analogy to classical mechanics, the Newton-Nelson
law was later shown to follow from a Hamilton-like stochastic
variational principle \cite{Y,GM}. Other
versions of the variational principle have been proposed in
\cite{N2,BCZ,P1,ROS}.

\noindent
Consider the case of a
nonrelativistic particle of mass $m$. Let
$\{\psi(x,t); t_0\le t\le t_1\}$ be the
solution of the {\it Schr\"{o}dinger equation}
\begin{equation}\label{H7} \frac{\partial{\psi}}{\partial{t}} =
\frac{i\hbar}{2m}\Delta\psi -
\frac{i}{\hbar}V(x)\psi, \end{equation}
such that
\begin{equation}\label{FA}||\nabla\psi||^2_2\in L^1_{{\rm loc}}[t_0,+\infty).
\end{equation}
This is Carlen's
{\em finite action condition}. Under these hypotheses, the Nelson measure
$P\in\D$
may be constructed on path space, \cite{C},\cite{Car}, \cite [Chapter
IV]{BCZ}, and references therein.
Namely, letting  $\Omega:={\cal C}([t_0,t_1],\R^n)$ the $n$-dimensional
continuous functions on $[t_0,t_1]$, under the  probability measure $P$,
the canonical coordinate process $x(t,\omega)=\omega(t)$ is an
$n$-dimensional Markov
diffusion process $\{x(t);t_0\le t\le t_1\}$,
called {\em Nelson's process}, having
(forward) Ito differential
\begin{equation}\label{N}
dx(t)=\left[\frac{\hbar}{m}\nabla \left(\Re
\log\psi(x(t),t) + \Im \log\psi(x(t),t)\right)\right]dt
+\sqrt{\frac{\hbar}{m}}dw(t),
\end{equation}
where $w$ is a standard, $n$-dimensional Wiener process. Moreover, the
probability density $\rho(\cdot,t)$ of
$x(t)$ satisfies
\begin{equation}\label{D}\rho(x,t)=|\psi(x,t)|^2,\quad \forall t \in [t_0,t_1].
\end{equation}
Following Nelson \cite{N1,N2}, for a finite-energy diffusion with
stochastic differentials (\ref{K1})-(\ref{K3}), we define the {\em
current} and {\em osmotic} drifts, respectively:
$$ v(t) = \frac{\beta (t) + \gamma (t)}{2},\quad		u(t) =
\frac{\beta (t) - \gamma (t)}{2}.
$$
Clearly $v$ is similar to the classical velocity, whereas $u$ is the
velocity due to the ``noise" which tends to zero when the diffusion
coefficient $\sigma^2$ tends to zero. In order to obtain a unique
time-reversal invariant differential \cite{P1}, we take a complex linear
combination of (\ref{K1})-(\ref{K3}), obtaining
\begin{eqnarray}\nonumber
&&x(t)-x(s)=\int_s^t\left[\frac{1-i}{2}\beta(\tau)+\frac{1+i}{2}\gamma
(\tau)\right]d\tau\\&&
+\frac{\sigma}{2}\left[(1-i)(w_+(t)-w_+(s))+(1+i)(w_-(t)-w_-(s))\right].
\nonumber\end{eqnarray}
Let us define the {\em quantum drift}
$$v_q(t):=\frac{1-i}{2}\beta(t)+\frac{1+i}{2}\gamma(t)=v(t)-iu(t),$$
and the {\it quantum noise}
$$w_q(t):=\frac{1-i}{2}w_+(t)+\frac{1+i}{2}w_-(t).
$$
Hence,
\begin{equation}\label{QSD}
x(t)-x(s)=\int_s^tv_q(\tau)d\tau+\sigma[w_q(t)-w_q(s)].
\end{equation}
This representation enjoys the
time reversal invariance property. It has been crucial in
order to develop a Lagrangian and a
Hamiltonian dynamics formalism in the context of Nelson's stochastic
mechanics in \cite{P1,P2,P3}.
Notice that replacing (\ref{K1})-(\ref{K3}) with (\ref{QSD}), we replace
the pair of real drifts $(v,u)$ by the unique {\em complex-valued} drift
$v-iu$ that tends correctly to $v$ when the diffusion coefficient tends
to zero.

\section{Quantum Schr\"{o}dinger bridges}\label{QSB}
We now  consider the same problem as at the end of Section \ref{SB}. We
have  a {\em reference stochastic model} $P\in \D$ given by the Nelson
measure on path space that has been constructed through a variational
principle
\cite{GM,N2,P1}. This Nelson process $x=\{x(t); t_0\le t\le t_1\}$ has an
associated solution $\{\psi(x,t):t_0\le t \le t_1\}$ of the
Schr\"{o}dinger equation in the sense that the quantum drift of $x$
is $v_q(t)=\frac{\hbar}{im}\nabla\log\psi(x(t),t)$ and the one-time
density of $x$ satisfies $\rho(x,t)=|\psi(x,t|^2$.  Suppose a position
measurement at time $t_1$ yields the probability density
$\rho_1(x)\neq |\psi(x,t_1)|^2$. We need a suitable variational
mechanism that, starting from $(P,\rho_1)$, produces the new
stochastic model in $\D(\rho_1)$. It is apparent that the variational
problem of Section \ref{SB} is not suitable as it preserves the backward
drift. Since in stochastic mechanics both differentials must be granted
the same status, we need to change both drifts as little as possible
given the new density $\rho_1$ at time $t_1$. Thus, we employ the
differential  (\ref{QSD}), and consider the variational problem:\\
\noindent
Extremize on
$(\tilde{x},\tilde{v}_q)\in(\D(\rho_1)\times\cal V)$ the functional
\begin{eqnarray}\nonumber J(\tilde{x},\tilde{v_q}):=
E\left\{\frac{1}{2}\log\frac{\tilde{\rho}_1(\tilde{x}(t_1))}
{\rho(\tilde{x}(t_1),t_1)}+\right.\\\left.\int_{t_1}^{t_2}
\frac{mi}{2\hbar}(v_q(\tilde{x}(t),t)-\tilde{v}_q(t))\cdot(v_q(\tilde{x}(t),t)
-\tilde{v}_q(t))\,dt\nonumber\right\}\nonumber\end{eqnarray}
subject to: $\tilde{x} \;{\rm has\;quantum\;drift\;(velocity)}\;
\tilde{v}_q.$\\
\noindent
Here  $v_q(x,t)=\frac{\hbar}{im}\nabla\log\psi(x,t)$ is quantum drift of
Nelson reference process, and $\D(\rho_1)$ is family
of finite-energy,
$\R^n$-valued diffusions on $[t_0,t_1]$ with diffusion coefficient
$\frac{\hbar}{m}$, and
having marginal $\rho_1$ at time $t_1$. Moreover, $\cal V$
denotes the family of finite-energy, $C^n$ - valued stochastic processes
on $[t_0,t_1]$. Following the same variational analysis as in \cite{P3},
we get a Hamilton-Jacobi-Bellman type equation
\begin{equation}\label{HJB}\frac{\partial{\varphi}}{\partial{t}}
+ v_q(x,t)\cdot \nabla \varphi(x,t)-\frac{i\hbar}{2m}\Delta
\varphi(x,t)=\frac{i\hbar}{2m}\nabla\varphi(x,t)\cdot\nabla\varphi(x,t),
\end{equation}
with terminal condition
$\varphi(x,t_1)=\frac{1}{2}\log\frac{\rho_1(x)}{\rho(x,t_1)}.$ Then
$\tilde{x}\in \D(\rho_1)$ with quantum drift
$$v_q(\tilde{x}(t),t)+\frac{\hbar}{mi}\nabla\varphi(\tilde{x}(t),t)$$
solves the extremization problem.
Write
$\psi(x,t_1)=\rho(x,t_1)^{\frac{1}{2}}\exp [\frac{i}{\hbar}S(x,t_1)]$, and let
$\{\tilde{\psi}(x,t)\}$ be solution of Schr\"{o}dinger equation
(\ref{H7}) on
$[t_0,t_1]$ with terminal condition
$$\tilde{\psi}(x,t_1)=
\rho_1(x)^{\frac{1}{2}}\exp [\frac{i}{\hbar}S(x,t_1)].$$
Next, notice that for $t\in[t_0,t_1]$
$$\left[\frac{\partial}{\partial
t}+v_q(x,t)\cdot\nabla-\frac{i\hbar}{2m}\Delta\right]
\left(\frac{\tilde{\psi}}{\psi}\right)=0,\;
\frac{\tilde{\psi}}{\psi}(x,t_1)=\left(\frac{\rho_1(x)}{\rho(x
,t_1)}\right)^{\frac{1}{2}},
$$
where $v_q(x,t)=\frac{\hbar}{im}\nabla\log\psi(x(t)$. It follows
that $\varphi(x,t):=\log\frac{\tilde{\psi}}{\psi}(x,t)$ solves
(\ref{HJB}), and the corresponding quantum drift is
$$v_q(\tilde{x}(t),t)+\frac{\hbar}{mi}\nabla\varphi(\tilde{x}(t),t)=
\frac{\hbar}{mi}\nabla
\log\tilde{\psi}(\tilde{x}(t),t).$$
Thus,  new process after
measurement at time $t_1$ ({\em quantum Schr\"{o}dinger bridge}) is just
the Nelson process associated to another solution $\tilde{\psi}$ of the
same Schr\"{o}dinger equation. Invariance of phase at $t_1$ follows
from the variational principle.

\section{Collapse of the wavefunction}
Consider the case where measurement at time $t_1$ only gives the
information that
$x$ lies in subset $D$ of configuration space $\R^n$ of the system. The
density
$\rho_1(x)$
just after measurement is
$$\rho_1(x)=\frac{\chi_D(x)\rho(x,t_1)}{\int_D \rho(x',t_1)dx'}\;,
$$
where $\rho(x,t_1)$ is density of  Nelson reference process $x$ at time
$t_1$.. Let $A$ be subspace of $L^2(\R^n)$ of functions with support in
$D$. Then $A^{\perp}$ is subspace of $L^2(\R^n)$ functions with
support in $D^c$. Decompose
$\psi(x,t_1)$ as
$$\psi(x,t_1)=\chi_D(x)\psi(x,t_1)+\chi_{D^c}(x)\psi(x,t_1)=\psi_1(x)+
\psi_2(x),
$$ with $\psi_1\in A$ and $\psi_2\in A^{\perp}$. The probability
$p_1$ of finding
particle  in $D$ is
$$p_1=\int_D |\psi(x,t_1)|^2dx=\int_{\R^n}|\psi_1(x)|^2dx.$$ If the result
of the measurement at time $t_1$ is that the particle lies in $D$,
the variational principle replaces
$\{x(t)\}$ with $\{\tilde{x}(t)\}$ and, consequently, replaces
$\psi(x,t_1)=\psi_1(x)+
\psi_2(x)$ with $\tilde{\psi}(x,t_1)$ where
$$\tilde{\psi}(x,t_1)=\rho(x)_1^{\frac{1}{2}}\exp
[\frac{i}{\hbar}S(x,t_1)]=
\frac{\psi_1(x)}{||\psi_1||_2}.$$
Postulating the variational principle of the previous section (rather
than the invariance of the phase at $t_1$), we have therefore obtained the
so-called ``collapse of the wavefunction", see e.g. \cite{Kel} and
references therein. The collapse is instantaneous, precisely as in the
orthodox theory. It occurs ``when the result of the measurement enters the
consciousness of the observer"
\cite{Wig}. We mention here that, outside of stochastic mechanics, there
exist alternative stochastic descriptions of (non instantaneous) quantum
state reduction such as those starting from a stochastic Schr\"{o}dinger
equation, see e.g. \cite{ADL} and references therein.
\section{Conclusion and outlook}
  We shall show elsewhere \cite{P4} that the
variational principle of section \ref{QSB} may be replaced by two
stochastic differential games with real velocities with an appealing
classical interpretation. We shall also show that, using Nelson's
observation in \cite {N2, N3} and this variational principle, it is
possible to obtain a completely satisfactory classical probabilistic
description of the two-slit experiment.

If the variational mechanism described here can be extended to the case
where both the initial and final quantum states are varied, it would
provide a general approach to the steering problem for quantum systems
(extending \cite{BFP}) that has important applications in quantum
computation
\cite{Ni}, control of molecular dynamics \cite{RVMK} and many other
fields.

\end{document}